\documentclass[reqno]{amsart}
\usepackage{amsmath,amssymb,cite}
\usepackage[mathscr]{euscript}

\theoremstyle{plain}
\newtheorem{theorem}{Theorem}[section]

\theoremstyle{definition}

\theoremstyle{remark}

\numberwithin{equation}{section}
\numberwithin{theorem}{section}

\def\be{\begin{equation}}
\def\ee{\end{equation}}
\def\bp{\begin{pmatrix}}
\def\ep{\end{pmatrix}}
\def\bea{\begin{eqnarray}}
\def\eea{\end{eqnarray}}

\def\\{\par\medskip}

\let\0=\noindent

\newcommand{\mc}[1]{{\mathcal #1}}

\newcommand{\bb}[1]{{\mathbb #1}}

\newcommand{\rme}{\mathrm{e}}

\newcommand{\rmd}{\mathrm{d}}

\title{Dynamics of infinite classical anharmonic crystals}

\author[P.\ Butt\`a]{Paolo Butt\`a}
\address{Paolo Butt\`a\hfill\break \indent
   Dipartimento di Matematica, 
   Universit\`a di Roma `La Sapienza' 
   \hfill\break \indent
   P.le Aldo Moro 5, 00185 Roma, Italy}
 \email{butta@mat.uniroma1.it}
\author[C.\ Marchioro]{Carlo Marchioro}
\address{Carlo Marchioro \hfill\break \indent
   Dipartimento di Matematica, 
   Universit\`a di Roma `La Sapienza' 
   \hfill\break \indent
   P.le Aldo Moro 5, 00185 Roma, Italy}
\email{marchior@mat.uniroma1.it}

\begin{document}

\begin{abstract}
We consider an unbounded lattice and at each point of this lattice an anharmonic oscillator, that interacts with its first neighborhoods via a pair potential $V$ and is subjected to a restoring force of potential $U$. We assume that $U$ and $V$ are even nonnegative polynomials of degree $2\sigma_1$ and $2\sigma_2$. We study the time evolution of this system, with a control of the growth in time of the local energy, and we give a nontrivial bound on the velocity of propagation of a perturbation. This is an extension to the case $\sigma_1 < 2\sigma_2-1$ of some already known results obtained for $\sigma_1 \geq 2\sigma_2-1$. 
\end{abstract}


\maketitle
\thispagestyle{empty}
 
\section{Introduction}
\label{sec:1}

In this paper we study the time evolution of an infinitely extended classical anharmonic system that represents a schematic model of a crystal. Given $m\in\bb N$ and $L_k\in \bb N\cup\{\infty\}$, $k=1,\ldots,m$, we consider the lattice $\Lambda=\{i=(i_1,\ldots,i_m)\in\bb Z^m \colon |i_k| \le L_k\; \forall\, k=1,\ldots,m\}$. At each point $i\in\Lambda$ there is a $\nu$-dimensional oscillator of unitary mass that interacts with its first neighborhoods and is subjected to a one-body force. More precisely, denoting by $(q_i,p_i)\in \bb R^\nu\times \bb R^\nu$ the position and momentum of the $i$-th oscillator, the system evolves via the classical Newtonian laws of motion with formal Hamiltonian,
\[
H(q,p)=\sum_{i\in\Lambda} \bigg\{ \frac{|p_i|^2}{2} + U(q_i) + \sum_{\substack{j\in\Lambda :|j-i|=1}} V(q_i-q_j)\bigg\}\;,
\]
where, for $\xi\in\bb R^\nu$, $U(\xi)$ and $V(\xi)$ are even nonnegative polynomials of the real variable $|\xi|$, of degree $2\sigma_1$ and $2\sigma_2$ (hence, with positive leading coefficients).

For finite $L_k$'s the dynamics is well defined by the theorem of existence and uniqueness for systems of ordinary differential equations. A different situation arises when some $L_k =\infty$. In this case, a correct choice of the phase space is crucial to obtain a well defined time evolution. Otherwise, there are reasonable initial data whose evolution is poorly defined, as shown by the following simple example of lack of uniqueness. Take $m=1$, $i\ge 1$ (a positive one-dimensional chain), and initial data $(q_i(0),p_i(0)) = (0,0)$ for any $i\ge 1$. Of course, the chain at rest for any time is the more natural solution, but we can exhibit another solution of the same Cauchy problem. We fix the motion of the first oscillator as $q_1(t) = \exp[-\lambda t^{-2}]$, $\lambda > 0$, and choose the motion $q_2(t)$ of the second oscillator in such a way that $\ddot q_1(t)$ evolves according to the equations of motion. We next choose the motion $q_3(t)$ of the third oscillator so that also $\ddot q_2(t)$ evolves according to the equations of motion. We can then proceed inductively, by obtaining an evolved state $\{(q_i(t),p_i(t))\}_{i\ge 1}$ solution to the equations of motion, vanishing for $t\to 0$ but otherwise different from zero; hence the uniqueness fails. The reason of this pathology is that this solution grows too fast as $i\to\infty$. (This counterexample is similar to the well known Tikhonov's example of lack of uniqueness for the heat equation).

The existence of solutions for such systems has been studied in \cite{LLL} for $\sigma_1\ge 2\sigma_2-1$ and, in general, in \cite{MPP2}. However, the results in these papers do not exclude that the initial data could change very much their local energy as time goes by. A better control of this growth, which has been proved in \cite{BCDM} in the case $\sigma_1\ge 2\sigma_2-1$, is the principal task of the present investigation. As we will see in the next section, this generalization needs a different approach to the problem. Our result, that we believe  interesting in itself, is a preliminary step for investigating the asymptotic in time behavior of the system. 

In this respect, in \cite{BCDM} a not trivial bound on the velocity of propagation of a perturbation has been given, provided that the system be in an equilibrium state. In the present case, in which we require only a positive $\sigma_1$, noticing that the existence of thermodynamic equilibria (i.e., the infinite volume Gibbs states) has been  proved in \cite{BMPP}, we are able to extend the bound of \cite{BCDM} to this more general situation. We refer to the next section for a precise formulation.

The system can become unbounded in one or more directions. We denote by $d$ the number of indexes $k$ such that $L_k=\infty$, and the results strongly depend on the value of $d$. Since the presence of $m-d$ directions along which the system is bounded is not relevant in the proofs, for the sake of notational simplicity we treat the case $m=d$, i.e., $\Lambda=\bb Z^d$.

There are possible generalizations of this result. We could change the geometry of the lattice, for instance, by considering a lattice in which each $L_k$, $k>1$, is a function of $i_1$ which diverges slowly as $i_1$ diverges. In this case, the system could behave as a crystal with a fractional $d$. Extensions to the case of not rectangular neither homogeneous lattices, as well as systems with finite or long-range interactions could be equally treated.

Finally, we remark that the time evolution of infinitely extended Hamiltonian systems has been investigated in several papers. For a short review on this topic, see for instance the Appendix of \cite{BCM5}.

\section{Notation and statement of the result}
\label{sec:2}

We consider a system of anharmonic $\nu$-dimensional oscillators in the $d$-dimensio\-nal lattice $\bb Z^d$. Denoting by $q_i\in\bb R^\nu$ and $p_i\in\bb R^\nu$ the position and momentum of the oscillator located at the lattice point $i\in \bb Z^d$, the mechanical state of the system is determined by the infinite sequence $x = \{x_i\}_{i\in \bb Z^d} = \{(q_i,p_i)\}_{i\in\bb Z^d}$ of positions and momenta of the oscillators. We shall denote by $\mc X$ the set of all such possible states, equipped with the product topology.

The time evolution $t\mapsto x(t) = \{(q_i(t),p_i(t))\}_{i\in\bb Z^d}$ is defined by the solutions of the following infinite set of coupled differential equations,
\begin{equation}
\label{p1}
\begin{cases} \dot q_i (t) = p_i(t)\;, \\
\dot p_i(t) = F_i(x(t))\;, \end{cases} \qquad i \in \bb Z^d\;,
\end{equation}
where the force $F_i(x)$ induced by the configuration $x=\{(q_i,p_i)\}_{i\in\bb Z^d}$ on the $i$-th oscillator is given by
\begin{equation}
\label{p2}
F_i(x) = - \nabla U(q_i) - \sum_{j:|j-i|=1} \nabla V(q_i-q_j)\;.
\end{equation}
As already detailed in the Introduction, for $\xi\in\bb R^\nu$, $V(\xi)$ and $U(\xi)$ are even nonnegative polynomials of the real variable $|\xi|$, with degree $2\sigma_1$ and $2\sigma_2$. Here $|i-j|$ is the distance between the points $i=(i_1,\ldots,i_d)$ and $j=(j_1,\ldots,j_d)$ defined by
\[
|i-j| = \sum_{\ell=1}^d |i_\ell-j_\ell|\;.
\]

Our task is to construct the dynamics for a class of initial conditions which is large enough to be typical for any reasonable thermodynamic equilibrium (or nonequilibrium) state, i.e., for any Borel probability measure $\omega$ on $\mc X$ that satisfies the following superstability estimate: there exists a positive constant $C_\omega$ such that, for any $\lambda$ small enough,
\begin{equation}
\label{p5}
\omega\big( \rme^{\lambda W_{\mu,k}}\big) \le \rme^{C_\omega (2k+1)^d} \qquad \forall\, \mu\in\bb Z^d \quad \forall\, k\in \bb N\;,
\end{equation}
with
\begin{equation}
\label{p3}
W_{\mu,k}(x) := \sum_{i\in\Lambda_{\mu,k}} \bigg\{ \frac{|p_i|^2}{2} + U(q_i) + 1 \bigg\} + \sum_{\substack{i,j\in\Lambda_{\mu,k}\\ |j-i|=1}} V(q_i-q_j)\;,
\end{equation}
where, for any integer $k\ge 1$, $\Lambda_{\mu,k}$ denotes the cube of center $\mu$ and side $2k+1$. 

To this purpose, it is enough to allow initial data with logarithmic divergences in the energy. More precisely, define
\begin{equation}
\label{p4}
Q(x) := \sup_{\mu\in\bb Z^d} \,\, \sup_{k>\log^{1/d}(e+|\mu|)} \,\, \frac{W_{\mu,k}(x)}{(2k+1)^d}
\end{equation}
and denote by $\mc X_0$ the following subset of $\mc X$,
\[
\mc X_0 = \{ x\in \mc X \, : \, Q(x)<\infty\}\;.
\] 
Then, for any $\lambda$ small enough (see, e.g., \cite{BCDM}),
\begin{equation}
\label{p6}
\lim_{N\to + \infty} e^{\lambda N} \, \omega(Q>N) = 0\;,
\end{equation}
which in particular yields $\omega(\mc X_0) = 1$. 

\begin{theorem}
\label{t:p1}
Let $\sigma=\max\{\sigma_1;\sigma_2\}$ and set $\eta := (\sigma-1)/\sigma$. If $\eta d<2$ there is a one-parameter group of transformations $\Phi_t \colon \mc X_0 \to \mc X_0$, $t\in \bb R$, such that $t \to \Phi_t(x)$ is the unique global solution to Eq.\ \eqref{p1} with initial condition $\Phi_0(x)=x$. Moreover, for any $\gamma \in (\eta d, 2)$ and $\beta>0$ there is a positive constant $C_{\gamma,\beta}$ such that, for any $t>0$, 
\begin{equation}
\label{p7}
Q(\Phi_t(x)) \le C_{\gamma,\beta}Q(x) \big[1+ t^{2d/(2-\gamma)}(1+ t^\beta) Q(x)^{\gamma/(2-\gamma)}\big] \;.
\end{equation}
\end{theorem}

The flow is obtained as the limit of an approximating sequence, defined via the time evolution of systems composed by a finite number of oscillators. We observe that the condition $\eta d < 2$ is always satisfied in dimension $d=1,2$, while it implies $\sigma<3$ in dimension $d=3$. Moreover, this condition can be removed if $\sigma_1\ge 2\sigma_2-1$. Indeed, this case can be treated as in \cite[Appendix]{BCDM} (where the specific choice $\sigma_1=2$ and $\sigma_2=1$ is detailed), via an a priori estimate on the growth in time of the energy density for these finite systems, which allows to perform the limit. It is worthwhile to mention that a priori energy estimates are the key ingredient also in several results concerning the dynamics of infinitely many particles moving in a continuum, see, e.g., \cite{BPY,BCCM,BCM3,BCM4,CCM,CMP,CMS,DF,FD}. Moreover, in the case of quasi-one-dimensional systems, such estimates turn out to be good enough to obtain not trivial results on the long time behavior, see, e.g., \cite{BCM1,BCM2,BMM,CM}.

In the general case under consideration, an a priori bound on the energy is not available. Indeed, the basic ingredient is an  integral inequality for the local energy, globally solvable in time and based on an estimate of the time derivative of the local energy in term of the local energy itself. In our context, this plan works only if $\sigma_1\ge 2\sigma_2-1$.\footnote{To see this, let $x^n(t)$ be the time evolution of the oscillators inside the cube $\Lambda_{0,n}$, assuming empty boundary conditions. Then, for any $\Lambda_{\mu,k}\subset \Lambda_{0,n}$ and recalling \eqref{p3},
\[
\frac{\rmd}{\rmd t} W_{\mu,k}(x^n(t)) \lesssim \underbrace{W_{\mu,k}(x^n(t))^{\frac 12}}_{\textrm{velocity}} \times \underbrace{W_{\mu,k+1}(x^n(t))^{\frac{2\sigma_2-1}{2\sigma}}}_{\mathrm{force}}\;,
\]
which can be shown to imply - we omit the details - that for $W_k^n(t) := \displaystyle \max_{\mu :\Lambda_{\mu,k}\subset\Lambda_{0,n}} W_{\mu,k}(x^n(t))$,
\[
W_k^n(t) \le W_k^n(0) + C \int_0^t\! W_k^n(s)^{\frac 12 + \frac{2\sigma_2-1}{2\sigma}} \rmd s\;.
\]
If $\frac 12 + \frac{2\sigma_2-1}{2\sigma} \le 1$, i.e., $\sigma_1\ge 2\sigma_2-1$, the above inequality can be solved globally in time, getting
\[
W_k^n(t) \le C(t) W_k^n(0) \le C(t) Q(x) \big[\log(\rme + n)+(2k+1)^d\big]\;,
\]
where we used \eqref{p4}, hence an a priori bound on $W_k^n(t)$, weakly depending on the size $n$ of the finite approximation.} To overcome this problem, we then adopt a different strategy, by constructing the flow as the unique limit of an infinite collection of approximating sequences, which differ among each other for the position where the finite systems of oscillators are located. This allows to approximate the time evolution of local quantities with their time evolution relative to a finite system of not too large size, and this in turn shows that the local energy of the system cannot change too drastically in time. We remark that there is a wide region of parameters in which both the present method and that of \cite{BCDM} are applicable and give slightly different bounds. Obviously, depending on the problem to face, one uses the most useful one.
 
As a corollary of Theorem \ref{t:p1}, we can extend to the general case the analysis done in \cite{BCDM}, which provides an upper bound on the velocity of disturbances at thermal equilibrium or, more precisely, an estimate of the size of time correlations. Indeed, the dynamical estimates of Theorem \ref{t:p1} can be significantly improved if the physical system is assumed at thermal equilibrium or, more generally, in any reasonable time invariant state. This allows to deduce not trivial bounds on the time correlations. Following \cite{BCDM,MPPT}, to quantify these correlations we use the Poisson brackets of local observables, classical version of the quantal brackets used in \cite{LR}.

To state precisely the result we need some preliminary definitions. Let $\mc U$ be the algebra of local observables, i.e., the set of functions $f\colon \mc X \to \bb R$ such that $f(x) = f_\Lambda(x_\Lambda)$ for some bounded set $\Lambda\subset \bb Z^d$ and some differentiable function $f_\Lambda$, depending on the finite set of real variables $x_\Lambda = \{x_i\}_{i\in\Lambda}$, which is differentiable with bounded derivative. If $f\in\mc U$ the function $\Phi_tf$ is defined by setting $\Phi_tf(x) = f(\Phi_t(x))$ for any $x\in\mc X_0$. The Poisson brackets $\{f,g\}$ of $f,g\in \mc U$ are defined by
\[
\{f,g\}(x) = \sum_{i\in\bb Z^d} \Big[(D_{q_i}f)^T D_{p_i}g -(D_{p_i}f)^T D_{q_i}g\Big] (x)\;.
\] 
Clearly, this definition makes sense also for nonlocal functions provided the series in the right-hand side converges. 

A time invariant state $\omega$ is a probability measure on $\mc X$ for which \eqref{p5} holds and such that $\omega(\Phi_t f) = \omega(f)$ for any $f\in\mc U$ and $t\in\bb R$. The typical example is given by the infinite Gibbs measure obtained as the thermodynamic limit with free boundary conditions. For the wide class of anharmonic systems considered here (i.e., without any condition between the degrees of the one-body potential $U$ and the interaction $V$), the existence of equilibrium states satisfying the superstability estimate \eqref{p5} has been proved in \cite{BMPP}. 

Given a pair of functions $f,g\colon\bb R^{2\nu} \to \bb R$, which are differentiable with bounded derivatives, we denote by $f_i$, resp.\ $g_j$, the local observable given by $f_i(x) = f(x_i)$, resp.\ $g_j(x) = g(x_j)$. 

\begin{theorem}
\label{t:p2}
Let $\omega$ be any time invariant state satisfying \eqref{p5}. Then, for each $f,g$ as above, $i\in \bb Z^d$, $\alpha > (4-\eta d)/(2-\eta d)$, and $b>0$, 
\begin{equation}
\label{p8}
\lim_{t\to\infty} \sup_{j:\,|i-j|> t \log^\alpha t} \rme^{bt} \, \{f_i,\Phi_t g_j\}(x) = 0\;,
\end{equation}
almost surely with respect to the probability measure $\omega$.
\end{theorem}

The meaning of Theorem \ref{t:p2} is clear by noticing that an improved version of \eqref{p8}, with $|i-j|> t \log^\alpha t$ replaced by $|i-j|> ct$ for some $c>0$, would imply a finite velocity of propagation. On the other hand, it is not clear to us that such improved estimate be valid for this type of observables. We rather expect  that a stronger decay of correlations should be true for suitable coarse grained versions of the local observables. 

With respect to \cite{BCDM}, the proof of Theorem \ref{t:p2} requires minor modifications but, for the sake of completeness, we give some details in Section \ref{sec:4}.

It is worthwhile to observe that Theorem \ref{t:p1} covers also the important case when the one-body interaction $U$ is absent. But one has to keep in mind that in this case the existence of states satisfying the superstability estimates is a delicate question. For example, in the harmonic case, there are not Gibbs measures in dimension $d=1,2$, and in dimension $d=3$ it is not known that they satisfy \eqref{p5}. 

We conclude the section with a notation warning: in the sequel, if not further specified, we shall denote by $C$ a generic positive constant whose numerical value may change from line to line and from one side to the other in an inequality; it may possibly depend only on the interactions $U$ and $V$ and on the dimensions $d$ and $\nu$.

\section{Proof of Theorem \ref{t:p1}}
\label{sec:3}

Fix an initial condition $x=\{x_i\}_{i\in\bb Z^d} = \{(q_i,p_i)\}_{i\in \bb Z^d}\in \mc X_0$. The solution to Eq.\ \eqref{p1} will be obtained as the limit of a suitable approximating sequence of finite dimensional evolutions, according to the following scheme. Given $\mu\in\bb Z^d$ and $n\in \bb N$, we call {\it $n$-partial dynamics around $\mu$} the flow $\Phi_t^{\mu,n}(x) = \big\{(q^{\mu,n}_i (t), p^{\mu,n}_i(t))\big\}_{i\in\bb Z^d}$ such that 
\be
\label{ap0}
(q^{\mu,n}_i (t), p^{\mu,n}_i(t)) = (q_i,p_i) \quad \forall\, i \in \bb Z^d\setminus \Lambda_{\mu,n}\;,
\ee
while $\big\{(q^{\mu,n}_i (t), p^{\mu,n}_i(t))\big\}_{i\in\Lambda_{\mu,n}}$ is the solution to the Cauchy problem,
\begin{equation}
\label{ap2}
\begin{cases} 
\dot q^{\mu,n}_i (t) = p^{\mu,n}_i(t)\;, \\ \dot p^{\mu,n}_i(t) = F^{\mu,n}_i(\Phi_t^{\mu,n}(x))\;, \\ (q^{\mu,n}_i (0), p^{\mu,n}_i(0)) = (q_i,p_i)\;, 
\end{cases} 
\end{equation}
where, for any $i\in\Lambda_{\mu,n}$,
\begin{equation}
\label{ap3}
F^{\mu,n}_i(x) =  - \nabla U(q_i) - \sum_{\substack{j\in\Lambda_{\mu,n}\\ |j-i|=1}} \nabla V(q_i-q_j)\;.
\end{equation}
We remark that the global existence and uniqueness of the solution to the problem \eqref{ap2} is guaranteed since $U$ and $V$ are bounded from below. 

The solution $\Phi_t(x)$ to Eq.\ \eqref{p1} will be obtained by setting
\begin{equation}
\label{ap1}
\Phi_t(x)_i = \lim_{n\to \infty} \Phi_t^{\mu,n}(x)_i \qquad \forall\, i\in\bb Z^d \;,
\end{equation}
provided that the limit in the right hand-side of \eqref{ap1} exists and does not depend on $\mu\in\bb Z^d$. 

To prove the existence of this limit, without loss of generality we consider the case $t>0$ and define,
\be
\label{ap3b}
\delta^{\mu,n}_i(t) = \max_{s\in [0,t]} \big|q^{\mu,n+1}_i (s) - q^{\mu,n}_i (s)\big|\;.
\ee

By \eqref{ap2}, for any $i\in\Lambda_{\mu,n}$,
\[
\delta^{\mu,n}_i(t) \le \int_0^t\! (t-s) \big| F^{\mu,n+1}_i(\Phi_s^{\mu,n+1}(x)) - F^{\mu,n}_i(\Phi_s^{\mu,n}(x)) \big| \,\rmd s\;.
\]
To estimate the difference of the forces, we use the following estimate, valid for any regular function $G\colon \bb R^\nu\to\bb R^\nu$,
\[
|G(\xi')-G(\xi)| = \bigg|\int_0^1\!  DG(\xi_\lambda) (\xi'-\xi)\,\rmd \lambda\bigg| \le |\xi'-\xi| \int_0^1\!  \|DG(\xi_\lambda)\| \,\rmd \lambda\;,
\]
where $\xi_\lambda := \lambda\xi'+(1-\lambda)\xi$, $DG$ is the Jacobian matrix of $G$, and $\|A\|$ denotes the operator norm of the matrix $A$. By \eqref{ap3} and \eqref{ap3b} we thus have, for any $i\in\Lambda_{\mu,n-1}$,
\be
\label{ap11}
\delta^{\mu,n}_i(t) \le  \int_0^t\! (t-s) \bigg[T^{\mu,n}_i(s) \delta^{\mu,n}_i(s) + \sum_{j: |j-i|=1}T^{\mu,n}_{i,j}(s) (\delta^{\mu,n}_i(s)+\delta^{\mu,n}_j(s))\bigg] \, \rmd s\;,
\ee
with
\[
\begin{split}
T^{\mu,n}_i(s) & = \int_0^1\! \big\| D^2U(\lambda q^{\mu,n+1}_i(s)+(1-\lambda) q^{\mu,n}_i(s))\big\| \,\rmd\lambda\;, \\ T^{\mu,n}_{i,j}(s) & = \int_0^1\! \big\| D^2V (\lambda (q^{\mu,n+1}_i(s)-q^{\mu,n+1}_j(s)) + (1-\lambda) (q^{\mu,n}_i(s)-q^{\mu,n}_j(s)))\big\|  \,\rmd \lambda\;,
\end{split}
\]
where $D^2U$ and $D^2V$ are the Hessian matrices of $U$ and $V$.

In view of the assumptions on the polynomials $U$ and $V$ and recalling \eqref{p3}, for any $i\in \Lambda_{\mu,n-1}$ and $j$ such that $|j-i|=1$ we have, 
\[
\begin{split}
T^{\mu,n}_i(s) &  \le C \big(1+|q^{\mu,n}_i (s)|^{2\sigma_1-2} + |q^{\mu,n+1}_i (s)|^{2\sigma_1-2} \big) \\ & \le C \big[W_{\mu,n}(\Phi_s^{\mu,n}(x))^{(\sigma_1-1)/\sigma_1} + W_{\mu,n+1}(\Phi_s^{\mu,n+1}(x))^{(\sigma_1-1)/\sigma_1}\big] \\ & \le C \big[W_{\mu,n}(\Phi_s^{\mu,n}(x))^\eta + W_{\mu,n+1}(\Phi_s^{\mu,n+1}(x))^\eta\big] \;, \\ T^{\mu,n}_{i,j}(s) & \le C\big[1+ |q^{\mu,n}_i(s)-q^{\mu,n}_j(s)|^{2\sigma_2-2} + |q^{\mu,n+1}_i(s)-q^{\mu,n+1}_j(s)|^{2\sigma_2-2}\big] \\ & \le C \big[W_{\mu,n}(\Phi_s^{\mu,n}(x))^{(\sigma_2-1)/\sigma_2} + W_{\mu,n+1}(\Phi_s^{\mu,n+1}(x))^{(\sigma_2-1)/\sigma_2}\big] \\ & \le C \big[W_{\mu,n}(\Phi_s^{\mu,n}(x))^\eta + W_{\mu,n+1}(\Phi_s^{\mu,n+1}(x))^\eta\big] \;,
\end{split}
\]
where we used the definitions $\sigma:=\max\{\sigma_1;\sigma_2\}$, $\eta := (\sigma-1)/\sigma$, and that $W_{\mu,n}\ge 1$.

Because of the energy conservation law, for any $n>1$ and $ s\in [0,t]$,
\be
\label{ap115}
W_{\mu,n}(\Phi_s^{\mu,n}(x)) = W_{\mu,n}(x)\;,  \quad W_{\mu,n+1}(\Phi_s^{\mu,n+1}(x)) = W_{\mu,n+1}(x)\;.
\ee
On the other hand, since $k\mapsto W_{\mu,k}(\cdot)$ is non decreasing and using \eqref{p4},
\be
\label{ap12}
\begin{split} 
W_{\mu,n}(x) & \le W_{\mu,n+1}(x) \le  W_{\mu,n+1+ \lfloor\log^{1/d}(\rme +|\mu|)\rfloor}(x) \le C\varphi_{\mu,n}(x)\;, \\ \mbox{with }& \varphi_{\mu,n}(x) :=  Q(x)  \big[ (2n+3)^d +\log(\rme+|\mu|) \big] \;,
\end{split}
\ee
where $\lfloor a \rfloor$ denotes the integer part of the positive number $a$. Plugging the above bounds in the right-hand side of \eqref{ap11} and taking the maximum for $i\in \Lambda_{\mu,k}$ with $k< n$, we finally obtain the integral inequality,
\[
u^{\mu,n}_k(t) \le C\, \varphi_{\mu,n}(x)^\eta \int_0^t\!(t-s) u^{\mu,n}_{k+1}(s) \,\rmd s\;,
\]
where
\[
u^{\mu,n}_k(t) :=  \max_{i\in \Lambda_{\mu,k}}\, \delta^{\mu,n}_i(t)\;, \quad k\le n\;.
\]
This inequality can be solved by iteration getting, for any $n>k$,
\[
u^{\mu,n}_k(t) \le \frac{\big[C \varphi_{\mu,n}(x)^\eta t^2\big]^{n-k}}{[2(n-k)]!} \, d^\mu_n(t)\;,
\]
where
\[
d^\mu_n(t)= \max_{s \in[0,t] }\, \max_{i\in\Lambda_{\mu,n}}\, \big( \big|q^{\mu,n+1}_i(s) - q_i\big| +\big|q^{\mu,n}_i(s) - q_i\big|\big) \;.
\]

In view of \eqref{p3}, \eqref{ap115}, and \eqref{ap12},
\[
d^\mu_n(t) \le t \,  \max_{s \in[0,t] }\, \max_{i\in\Lambda_{\mu,n}} \big\{ \big|p^{\mu,n+1}_i(s) \big| +  \big|p^{\mu,n}_i(s) \big| \big\} \le  Ct\, \varphi_{\mu,n}(x)^{\frac 12}\;,
\]
hence, by Stirling formula, 
\be
\label{uk}
u^{\mu,n}_k(t) \le \frac{C^{n-k} \,t^{2(n-k)+1} \, \varphi_{\mu,n}(x)^{\eta(n-k)+\frac 12}}{(n-k)^{2(n-k)}}\;.
\ee

Given $\gamma\in (\eta d,2)$ and $\beta>0$ as in the statement of the theorem, we claim that if 
\be
\label{nk}
n_k^* = 2k + \big\lfloor A\big(1+t^2(1+t^{\beta'})Q(x)^{\gamma/d}\big)^{1/(2-\gamma)} \log^{1/d}(\rme+|\mu|)\big\rfloor
\ee
with $A>1$ large enough and $\beta':=(2-\gamma)\beta/d$, then
\be
\label{nk1}
u^{\mu,n}_k(t) \le 2^{-(n-k)}\qquad \forall\, n\ge n_k^*\;.
\ee
To prove the claim, we first observe that if $n\ge n_k^*$ then $n\le 2(n-k)$, so that 
\[
\varphi_{\mu,n}(x)^{\eta(n-k)+\frac 12} \le C^{n-k} \big\{Q(x)[(n-k)^d +\log(\rme+|\mu|])\big\}^{\eta(n-k)+\frac 12}\qquad \forall\, n\ge n_k^*\;.
\]
Moreover, since $n_k^*-k\ge A$, by choosing $A$ large enough we have $\eta(n-k)+\frac 12 \le \gamma (n-k)/d$ and $2(n-k)+1\le (2+\beta')(n-k)$ for any $n\ge n_k^*$. In particular, $t^{2(n-k)+1}\le [t^2(1+t^{\beta'})]^{n-k}$ for any $t>0$ and $n\ge n_k^*$. Therefore, the estimate \eqref{nk1} follows provided that, given $C>0$, we can find $A$ large enough such that, 
\[
C t^2(1+t^{\beta'}) Q(x)^{\gamma/d} \, \frac{(n-k)^\gamma + \log^{\gamma/d}(\rme+|\mu|)}{(n-k)^2} \le \frac 12 \qquad \forall\, n\ge n_k^*\;.
\]
But this is true, since in view of the definition \eqref{nk}, 
\[
n-k\ge A\big(1+t^2(1+t^{\beta'})Q(x)^{\gamma/d}\big)^{1/(2-\gamma)} \log^{1/d}(\rme+|\mu|)\qquad \forall\, n\ge n_k^*\;,
\]
so that
\[
\begin{split}
t^2(1+t^{\beta'}) &  Q(x)^{\gamma/d} \, \frac{(n-k)^\gamma + \log^{\gamma/d}(\rme+|\mu|)}{(n-k)^2} = \frac{t^2(1+t^{\beta'}) Q(x)^{\gamma/d} }{(n-k)^{2-\gamma}} \\ & \quad + \frac{t^2(1+t^{\beta'}) Q(x)^{\gamma/d}\log^{\gamma/d}(\rme+|\mu|)}{(n-k)^2} \\ & \le \frac{1}{A^{2-\gamma}} + \frac{t^2(1+t^{\beta'})Q(x)^{\gamma/d}}{A^2[1+t^2(1+t^{\beta'})Q(x)^{\gamma/d}]^{2/(2-\gamma)}} \log^{-\frac{2-\gamma}d}(\rme+|\mu|) \\ & \le \frac{1}{A^{2-\gamma}} + \frac{1}{A^2}\qquad \forall\, n\ge n_k^*\;.
\end{split}
\]

By \eqref{nk1} we get that $u^{\mu,n}_k(t)$ is $n$-summable, which implies the existence of $q_i(t) := \lim_{n\to\infty}q^{\mu,n}_i(t)$ for any fixed $i\in \bb Z^d$. To show the convergence of the corresponding momentum $p^{\mu,n}_i(t)$ to some function $p_i(t)$ it is enough to take the limit $n\to\infty$ in both sides of the identity,
\[
p^{\mu,n}_i(t) = p_i + \int_0^t\! \Big[\nabla U(q^{\mu,n}_i(s)) + \sum_{\substack{j\in\Lambda_{\mu,n}\\ |j-i|=1}} \nabla V(q^{\mu,n}_i(s)-q^{\mu,n}_j(s))\Big] \, \rmd s\;,
\] 
which in turn shows not only the existence of the limit in the right-hand side of \eqref{ap1}, but also that it is solution to \eqref{p1}. 

It remains to prove that the limit $\Phi_t(x)$ in \eqref{ap1} is independent of the choice of $\mu$. To this purpose, fix $\mu,\mu'\in\bb Z^d$ and define, for any $n$ large enough to have $\Lambda_{\mu,n}\cap\Lambda_{\mu',n} \ne\emptyset$,
\[
\delta^{\mu,\mu'}_{i,n}(t) := \max_{s\in [0,t]} \big|q^{\mu,n}_i(s) - q^{\mu',n}_i(s)\big|\;, \quad i\in \Lambda_{\mu,n}\cap\Lambda_{\mu',n}\;.
\]
By \eqref{ap2} and \eqref{ap3}, for any $i\in \Lambda_{\mu,n-1}\cap\Lambda_{\mu',n-1}$ we have (compare with \eqref{ap11}),
\[
\begin{split}
\delta^{\mu,\mu'}_{i,n}(t) & \le \int_0^t\! (t-s) \big| F^{\mu',n}_i(\Phi_s^{\mu',n}(x)) - F^{\mu,n}_i(\Phi_s^{\mu,n}(x)) \big| \,\rmd s  \\ & \le \int_0^t\! (t-s) T^{\mu,\mu'}_{i,n}(s) \delta^{\mu,\mu'}_{i,n}(s) \, \rmd s + \sum_{j: |j-i|=1} \int_0^t\! (t-s) T^{\mu,\mu'}_{i,j,n}(s) \delta^{\mu,\mu'}_{j,n}(s) \, \rmd s\;,
\end{split}
\]
where
\[
\begin{split}
T^{\mu,\mu'}_{i,n}(s) & = \int_0^1\! \big\| D^2U(\lambda q^{\mu',n}_i(s)+(1-\lambda) q^{\mu,n}_i(s))\big\| \,\rmd\lambda\;, \\ T^{\mu,\mu'}_{i,j,n}(s) & = \int_0^1\! \big\| D^2V(\lambda (q^{\mu',n}_i(s)-q^{\mu',n}_j(s)) + (1-\lambda) (q^{\mu,n}_i(s)-q^{\mu,n}_j(s)))\big\|  \,\rmd \lambda\;,
\end{split}
\]
that, by conservation of energy and using \eqref{p4}, are bounded by $C \big[W_{\mu,n}(x)^\eta + W_{\mu',n}(x)^\eta\big]$.
We then apply the same iterative argument used to prove \eqref{uk} to get
\[
\delta^{\mu,\mu'}_{i,n}(t)  \le \frac{C^{k_i(n)} \,t^{2k_i(n)} \, \varphi^{\mu,\mu'}_n(x)^{\eta k_i(n)+\frac 12}}{k_i(n)^{2k_i(n)}}\;,
\]
where $\varphi^{\mu,\mu'}_n(x) := Q(x)  \big\{ (2n+1)^d +\log[(\rme+|\mu|) (\rme+|\mu'|)] \big\}$ and $k_i(n)$ is the largest integer $k$ such that $\Lambda_{i,k},\subset \Lambda_{\mu,n-1}\cap\Lambda_{\mu',n-1}$. Since $k_i(n) = n + o(1)$, the previous bound implies that $\tilde \delta^{\mu,\mu'}_{i,n}(t) \to 0$ as $n\to+\infty$ for any choice of $i,\mu,\mu'\in\bb Z^d$ and $t\in\bb R$. 

To control the growth in time of the energy, we first deduce a bound on the rate of convergence of the partial dynamics to its limit.

By \eqref{nk1}, for some constant $c_1 >0$,
\begin{equation}
\label{ap5}
\max_{i\in \Lambda_{\mu,k}} \max_{s\in [0,t]} |q_i(s) - q^{\mu,n_k^*}_i(s)| \le \sum_{n\ge n_k^*} u^{\mu,n}_k(t)  \le  \rme^{-c_1 n_k^*}\;.
\end{equation}
Concerning the momenta, in view of \eqref{p2} and \eqref{ap3}, for any $i\in \Lambda_{\mu,k}$ we have, setting $\delta q^{\mu,n_k^*}_i(s) = |q^{\mu,n_k^*}_i(s) - q_i(s)|$, 
\[
\begin{split}
& \max_{s\in [0,t]}|p_i (s) - p^{\mu,n_k^*}_i (s)| \le \int_0^t\! \big| F_i(\Phi_s(x)) - F^{\mu,n_k^*}_i(\Phi_s^{\mu,n_k^*}(x)) \big| \,\rmd s \\ &  \le C \int_0^t\! \bigg\{\big(1+|q^{\mu,n_k^*}_i(s)|^{2\sigma_1-2}\big)\,\delta q^{\mu,n_k^*}_i(s) \\ &  \quad + \sum_{j:|j-i|=1} \big(1+ |q^{\mu,n_k^*}_i(s)-q^{\mu,n_k^*}_j(s)|^{2\sigma_2-2}\big)\, \big(\delta q^{\mu,n_k^*}_i (s)+\delta q^{\mu,n_k^*}_j(s)\big)\bigg\} \,\rmd s\;,
\end{split}
\]
where, in estimating the Lipschitz constant of the force, we used that by \eqref{ap5} the Hessian norms of $\|D^2U\|$ and $\|D^2V\|$ are computed in points close to $q^{\mu,n}_i(s)$ and $q^{\mu,n}_i(s)-q^{\mu,n}_j(s)$. By the last estimate, using again \eqref{ap5} and the definition \eqref{p3}, we thus get,
\[
\max_{s\in [0,t]}|p_i (s) - p^{\mu,n_k^*}_i (s)|  \le  C t\,  W_{\mu,k+1}(\Phi_s^{\mu,n_k^*}(x))^\eta\,  \rme^{-c_1 n_k^*} \;.
\]
On the other hand, 
\be
\label{ap7}
W_{\mu,k+1}\big(\Phi_t^{\mu,n_k^*}(x)\big) \le W_{\mu,n_k^*}\big(\Phi_t^{\mu,n_k^*}(x)\big) = W_{\mu,n_k^*}(x) \le Q(x)(2n_k^*+1)^d\;,
\ee
where the first inequality follows since $k\mapsto W_{\mu,k}(\cdot)$ is non decreasing and $n_k^*>k+1$, the successive equality by energy conservation, and the last inequality since $n_k^*>\log^{1/d}(\rme+|\mu|)$ (recall that in \eqref{nk} we assume $A>1$). In conclusion, there is a constant $c >0$ such that, for any $\mu\in \bb Z^d$ and $k\in\bb N$,
\begin{equation}
\label{ap5b}
\max_{i\in \Lambda_{\mu,k}} \max_{s\in [0,t]} |\Phi^{\mu,n_k^*}_s(x)_i - \Phi_s(x)_i| \le  \rme^{-c n_k^*}\;.
\end{equation}

We are now in position to prove the estimate \eqref{p7}, which in particular shows that $\Phi_t(x)\in\mc X_0$ for any $t\in\bb R$. We have, 
\[
\begin{split}
W_{\mu,k}\big(\Phi_t(x)\big) & \le W_{\mu,k}\big(\Phi_t^{\mu,n_k^*}(x)\big) + \big| W_{\mu,k}\big( \Phi_t(x)\big) - W_{\mu,k}\big( \Phi_t^{\mu,n_k^*}(x)\big)\big| \\ & \le W_{\mu,k}\big(\Phi_t^{\mu,n_k^*}(x)\big) + C\, W_{\mu,k}\big(\Phi_t^{\mu,n_k^*}(x)\big)^{(2\sigma-1)/(2\sigma)}  \rme^{-c n_k^*} \\ & \le Q(x)(2n_k^*+1)^d + \big[Q(x)(2n_k^*+1)^d\big]^{(2\sigma-1)/(2\sigma)} \rme^{-c n_k^*}\;,
\end{split}
\]
where we used \eqref{ap7} and that the Lipschitz constant of the energy is computed in points close to $\Phi_t^{\mu,n_k^*}(x)$. By \eqref{nk} and using that $Q(x)\ge 1$, see \eqref{p3} and \eqref{p4}, the last display implies that if $k>\log^{1/d}(\rme + |\mu|)$ then 
\[
\begin{split}
\frac{W_{\mu,k}\big(\Phi_t(x)\big)}{(2k+1)^d} & \le Q(x) \Big[2+A\big(1+t^2(1+t^{\beta'})Q(x)^{\gamma/d}\big)^{1/(2-\gamma)}\Big]^d + C Q(x)^{(2\sigma-1)/(2\sigma)} \\ & \le C Q(x) \big[1+A^d+t^{2d/(2-\gamma)}(1+t^{\beta'd/(2-\gamma)}) Q(x)^{\gamma/(2-\gamma)}\big]\;.
\end{split}
\]
Therefore, as $\beta':=(2-\gamma)\beta/d$, the ratio $W_{\mu,k}\big(\Phi_t(x)\big)/(2k+1)^d$ is bounded from above by the right-hand side of \eqref{p7} for a suitable positive constant $C_{\gamma,\beta}$. The theorem is thus proved.

\section{Proof of Theorem \ref{t:p2}}
\label{sec:4}

In this section we prove Theorem \ref{t:p2}. The estimate \eqref{p7} is too bad to prove \eqref{p8} for any $x\in \mc X_0$. Following \cite{BCDM}, we instead take advantage of the time invariance of the state $\omega$ to construct a smaller set $\mc B\subset \mc X_0$ of full $\omega$-measure, in which the growth of the local energy is under control for all sufficiently large integer times $k\in\bb N$. The estimate \eqref{p7} can be then fruitfully used to bound the energy during the time intervals $[k,k+1]$. As a result, we obtain a control on $Q(\Phi_t(x))$ for any $x\in \mc B$ and $t\ge 0$, which is good enough to handle the correlations.

Given $\alpha>(4-\eta d)/(2-\eta d)$ as in the statement of Theorem \ref{t:p2}, we fix $\gamma \in (\eta d,2)$ such that $\alpha>(4-\gamma)/(2-\gamma)$ and $\delta$ such that
\be
\label{del}
1 < \delta < \Big(\frac\alpha\eta-\frac 12\Big) (2-\gamma)\;,
\ee
where we remark that the right-hand side is larger than $1$ since, in view of the choices of $\gamma$ and $\alpha$,
\[
\Big(\frac\alpha\eta-\frac 12\Big) (2-\gamma) > \Big(\frac d\gamma \frac{4-\gamma}{2-\gamma} - \frac 12\Big) (2-\gamma) = 1 + \frac{(2d-\gamma)(4-\gamma)}{2\gamma}>1\;.
\]
We define, 
\[
\mc B := \bigcup_{n=1}^\infty \bigcap_{k=n}^\infty \mc B_k\;, \qquad \mc B_k := \big\{x\in\mc X \colon Q(\Phi_k(x)) \le \log^\delta k \big\}\;.
\]
Setting $\mc B_k^\complement := \mc X \setminus \mc B_k$, since $\omega$ is time invariant, 
\[
\omega\big(\mc B_k^\complement \big) = \omega\big(Q \circ \Phi_k > \log^\delta k \big) = \omega \big(Q > \log^\delta k \big)\;.
\]
It follows, by \eqref{p6}, that if $\lambda$ is small enough then $\rme^{\lambda\log^\delta k}\omega\big(\mc B_k^\complement \big) \to 0$ as $k \to \infty$. In particular, since $\delta>1$, $\sum_k\omega \big(\mc B_k^\complement \big) < \infty$, whence $\omega(\mc B)=1$ by the Borel-Cantelli lemma.
 
Recalling that $(q_j(t),p_j(t))$ denotes the $j$-th coordinates of $\Phi_t(x)$, we introduce the $2\nu\times 2\nu$ Jacobian matrix given by 
\begin{equation}
\label{p10}
\Delta_{j,i}(t,x) = D_{x_i} \Phi_t(x)_j = \begin{pmatrix} D_{q_i} q_j(t) &  D_{p_i} q_j(t) \\ D_{q_i} p_j(t) & D_{p_i} p_j(t) \end{pmatrix}\;.
\end{equation}
It is straightforward to verify that $|\{f_i,\Phi_t g_j\}(x) |\le C\|Df\|_\infty\|Dg\|_\infty \big\| \Delta_{j,i}(t,x)\big\| $. Therefore, since $\omega(\mc B) = 1$, Theorem \ref{t:p2} follows once we show that, for any $b>0$,
\begin{equation}
\label{p11}
\lim_{t\to\infty} \sup_{j\,:\,|i-j|> t \log^\alpha t} e^{bt}\, 
\big\| \Delta_{j,i}(t,x)\big\| = 0 \qquad \forall\, x\in\mc B\;. 
\end{equation}

By \eqref{p1}, the trajectory $(q_j(t),p_j(t))$ satisfies the equation,
\[
\begin{pmatrix} q_j(t) \\ p_j(t) \end{pmatrix} = \begin{pmatrix} q_j + p_j t \\ p_j \end{pmatrix} + \int_0^t\!\begin{pmatrix} (t-s) F_j(\Phi_s(x)) \\ F_j(\Phi_s(x))\end{pmatrix}\, \rmd s \;, 
\]
from which we get, recalling \eqref{p2},
\begin{equation}
\label{p14}
\Delta_{j,i}(t,x) = \begin{pmatrix}1 & t \\ 0 & 1 \end{pmatrix} \delta_{j,i} + \sum_{h\in\bb Z^d} \int_0^t\! B_{j,h}(s) \begin{pmatrix}t-s & 0 \\ 1 & 0 \end{pmatrix} \Delta_{h,i}(s,x)\, \rmd s\;, 
\end{equation}
with
\begin{equation}
\label{p15}
B_{j,h}(s) = - D^2 U(q_j(s)) \delta_{j,h} - \sum_{\ell : |\ell-j|=1} D^2V(q_j(s)-q_\ell(s)) (\delta_{j,h} -\delta_{\ell, h}) \;.
\end{equation}
The integral equation \eqref{p14} can be solved by iteration, getting
\begin{equation}
\label{p16}
\begin{split}
\Delta_{j,i}(t,x) & = \begin{pmatrix}1 & t \\ 0 & 1 \end{pmatrix} \delta_{j,i} + \sum_{n=1}^\infty \int_0^t\!
\int_0^{t_1}\! \cdots \int_0^{t_{n-1}}\! G_j(t_1,\ldots,t_n) \\ & \times \, \begin{pmatrix}(t-t_1)(t_1-t_2)\cdots(t_{n-1}-t_n) & 0 \\ (t_1-t_2)\cdots(t_{n-1}-t_n) & 0 \end{pmatrix} \begin{pmatrix}1 & t_n \\ 0 & 1 \end{pmatrix} \, \rmd t_n\cdots\rmd t_2\, \rmd t_1\;,
\end{split}
\end{equation}
where
\begin{equation}
\label{p16bis}
G_j(t_1,\ldots,t_n) = \begin{cases}
{\displaystyle \sum_{k_1,\ldots,k_n}} B_{j,k_1}(t_1) \cdots
B_{k_n,i}(t_n) & \text{ if }\, |j-i|\le n, \\ 0 & \text{ otherwise}.
\end{cases}
\end{equation}

From \eqref{p15}, by using \eqref{p3}, \eqref{p4}, and the hypothesis on $U$ and $V$, it follows that
\begin{equation}
\label{p17}
\big\|B_{k,h}(s)\big\| \le C \big[Q(\Phi_s(x)) \log(\rme+|k|)\big]^{\eta} \qquad \forall\, k,h\in\bb Z^d \quad \forall\, s\in\bb R\;. 
\end{equation}
The sum in \eqref{p16bis} involves only sites $k_\ell$ such that $|k_\ell| \le |i|+n$. Since the number of $n$-step walks, starting from a given site and with the possible presence of some permanences, is bounded by $(2d+1)^n$, by \eqref{p16} and \eqref{p17} we have,
\begin{equation}
\label{p12}
\begin{split}
\big\|\Delta_{j,i}(t,x)\big\| & \le  (1+t) \sum_{n=|j-i|}^\infty \frac1{(2n)!}\Big[C t^2  \log^{\eta}(\rme+|i|+n)\sup_{0\le s\le t} Q(\Phi_s(x))^\eta\Big]^n \\ & \le  (1+t) \sum_{n=|j-i|}^\infty \bigg[\frac{C t^2\log^{\eta}(\rme+|i|+n)}{n^2} \sup_{0\le s\le t} Q(\Phi_s(x))^\eta\bigg]^n\;,
\end{split}
\end{equation}
where we used the Stirling formula in the last inequality.

We can now prove \eqref{p11}. We fix $x\in \mc B$ and observe that, by the definition of $\mc B$, there exists a positive integer $n_0 = n_0(x)$ such that $Q(\Phi_k(x)) \le \log^\delta k$ for any $k\ge n_0$.  Using \eqref{p7}, for any $t>n_0$, 
\[
\begin{split}
\sup_{0\le s\le t} Q(\Phi_s(x)) & \le  \sup_{0\le s \le n_0} Q(\Phi_s(x)) + \max_{k=n_0,\ldots, [t]}\: \sup_{0\le s\le 1} Q\big(\Phi_s\big(\Phi_{k}(x)\big)\big) \\ & \le C_{\gamma,\beta} Q(x) \big[1+ n_0^{2d/(2-\gamma)} (1+n_0^\beta) Q(x)^{\gamma/(2-\gamma)} \big] \\ & \quad + \, \max_{k=n_0,\ldots, [t]}\: C_{\gamma,\beta} Q(\Phi_{k}(x)) \big[1+ 2Q(\Phi_{k}(x))^{\gamma/(2-\gamma)} \big]\;,
\end{split}
\]
whence, for $t_0=t_0(x)$ sufficiently large,
\[
\sup_{0\le s\le t} Q(\Phi_s(x)) \le  3C_{\gamma,\beta} (\log t)^{2\delta/(2-\gamma)} \qquad \forall\, t\ge t_0\;.
\]
By \eqref{p12} we thus obtain that, for any $t$ large enough,
\be
\label{fd}
\big\|\Delta_{j,i}(t,x)\big\|  \le (1+t) \sum_{n= |j-i|}^\infty \bigg[\frac{C C_{\gamma,\beta} t^2 \log^{\eta}(\rme+|i|+n) (\log t) ^{2\delta\eta/(2-\gamma)} }{n^2}  \bigg]^n\;.
\ee
Since the parameter $\delta$ satisfies the upper bound in \eqref{del}, the ratio in the square brackets in the series vanishes as $t\to\infty$ for any $n\ge t\log^\alpha t$. Therefore, we can find $t_1\ge t_0$ such that if $|j-i|>t\log^\alpha t$ then the right-hand side in \eqref{fd} is bounded by $C\exp\big\{-t\log^\alpha t\big\}$ for any $t\ge t_1$. The limit \eqref{p11} is thus proved.

\subsection*{Acknowledgments}
Work performed under the auspices of the Italian Ministry of the University (MIUR).


\begin{thebibliography}{99}

\small

\bibitem{BPY} C. Bahan, Y.M. Park and H.J. Yoo, Non equilibrium dynamics of infinite particle systems with infinite range interaction, {\em J. Math. Phys. \bf 40}:4337--4358 (1999).

\bibitem{BMPP} G. Benfatto, C. Marchioro, E. Presutti and M. Pulvirenti, Superstability estimates for anharmonic systems, {\em J. Stat. Phys. \bf 22}:349--362 (1980).

\bibitem{BCM1} P. Butt\`a, E. Caglioti and C. Marchioro, On the motion of a charged particle interacting with an infinitely extended system, {\em Comm. Math. Phys. \bf 233}:545--569 (2003).

\bibitem{BCM2} P. Butt\`a, E. Caglioti and C. Marchioro, On the violation of Ohm's law for bounded interactions: a one dimensional system, {\em Comm. Math. Phys. \bf 249}:353--382 (2004).

\bibitem{BCM3}  P. Butt\`a, G. Cavallaro and C. Marchioro, Time evolution of two dimensional systems with infinitely many particles mutually interacting via very singular forces, {\em J. Stat. Phys. \bf 147}:412--423 (2012).

\bibitem{BCM4}  P. Butt\`a, G. Cavallaro and C. Marchioro, Dynamics of infinitely extended hard core systems, {\em Rep. Math. Phys. \bf 72}:369--377 (2013).

\bibitem{BCM5}  P. Butt\`a, G. Cavallaro and C. Marchioro, {\em Mathematical models of viscous friction}, Lecture Notes in Mathematics, \textbf{2135}, Springer, Cham, 2015.

\bibitem{BCDM} P. Butt\`a, E. Caglioti, S. Di Ruzza, and C. Marchioro,  On the propagation of a perturbation in an anharmonic system, {\em J. Stat. Phys. \bf 127} 313--325 (2007).
 
\bibitem{BCCM} P. Butt\`a, S. Caprino, G. Cavallaro and C. Marchioro, On the dynamics of infinitely many particles with magnetic confinement, {\em Boll. Unione Mat. Ital. Sez. B Artic. Ric. Mat. (8) \bf 9}:371--395 (2006). 

\bibitem{BMM} P. Butt\`a, F. Manzo and C. Marchioro, A simple Hamiltonian model of runaway particle with singular interaction, {\em Math. Mod. Meth. In Appl. Sciences \bf 15}:753--766 (2005).

\bibitem{CM} E. Caglioti and C. Marchioro, On the long time behavior of a particle in an infinitely extended
system in one dimension, {\em J. Stat. Phys. \bf 106}:663--680 (2002).

\bibitem{CMP} E. Caglioti, C. Marchioro and M. Pulvirenti, Non-equilibrium dynamics of three-dimensional infinite particle systems, {\em Comm. Math. Phys. \bf 215}:25--43 (2000).

\bibitem{CCM} S. Caprino, G. Cavallaro, C. Marchioro, Time evolution  of an
infinitely extended  Vlasov fluid with singular mutual interactions,
{\em J. Stat. Phys.},  published online (2015). 

\bibitem{CMS} G. Cavallaro, C. Marchioro and C. Spitoni, Dynamics of infinitely many particles mutually interacting in three dimensions via a bounded superstable long-range potential, {\em J. Stat. Phys. \bf 120}:367--416 (2005).

\bibitem{DF} R.L. Dobrushin and J. Fritz, Non equilibrium dynamics of one-dimensional infinite particle system with hard-core interaction, {\em Comm. Math. Phys. \bf 55}:275--292 (1977).

\bibitem{FD} J. Fritz and R.L. Dobrushin, Non-equilibrium dynamics of two-dimensional infinite particle systems with 
a singular interaction, {\em Comm. Math. Phys. \bf 57}:67--81 (1977).

\bibitem{LLL} O.E. Lanford, J.L. Lebowitz and H. Lieb, Time evolution of infinite anharmonic systems, {\em J. Stat. Phys. \bf 16}:453--461 (1977).

\bibitem{LR} E. Lieb and D.W. Robinson, The finite group velocity of quantum spin systems, {\em  Comm. Math. Phys. \bf 28}:251--257 (1972).

\bibitem{MPP2} C. Marchioro, A. Pellegrinotti and M. Pulvirenti, On the dynamics of infinite anharmonic systems, {\em J. Math. Phys. \bf 22}:1740--1745 (1981).

\bibitem{MPPT} C. Marchioro, A. Pellegrinotti, M. Pulvirenti  and L. Triolo, Velocity of a perturbation in infinite lattice systems, {\em J. Stat. Phys. \bf 19}:499--510 (1978).

\end{thebibliography}
\end{document}